# Gene-Environment Interaction in the Era of Precision Medicine – Fix the Potholes or Start Building a New Road?

*José M Álvarez-Castro*[1]

Department of Education, University and Professional Training, Xunta de Galicia, Santiago de Compostela, Galicia, CP 15781, Spain.

**Summary**

Genetic mapping sprung in the last decade of the 20$^{th}$ century with the development of statistical procedures putting classical models of genetic effects together with molecular biology techniques. It eventually became clear that those models, originally developed to serve other purposes, implied limitations at different stages of the analyses—disclosing loci, measuring their effects and providing additional parameters for adequate biological/medical interpretations. The present paper is aimed to ponder whether it is realistic and worth to try and further amend classical models of genetic effects or it proves more sensible to undertake alternative theoretical strategies instead. In order to further feed into that debate, mathematical developments for gene-environment interaction stemming from the classical models of genetic effects are here revised and brought up-to-date with the prospects present-day available data bestow, particularly in the context of precision medicine. Those developments strengthen the methodology required to overcome the COVID-19 pandemic.

**Introduction**

Steps forward in scientific progress are often accompanied with expectations beyond objective appraisal. Quantitative trait locus experiments rose thirty years back and substantial resources were since soon after expectantly invested for elucidating genetic architectures of traits of economical importance (see e.g. Rifkin, 2012). In turn, the latest decade witnessed a swift switch of major efforts to aid livestock production and plant breeding towards genomic prediction (see e.g. Gondro et al., 2013). Initially developed for model species, genetic mapping of human traits became stunningly possible at the beginning of the current century by means of The International HapMap Project (International HapMap, 2003) and genome-wide association studies (GWAS; see e.g. Gondro et al., 2013) but, similarly, its potential for dissecting the genetic basis of diseases is already openly questioned nowadays (see e.g. Teperino, 2020a).

In this context, the first half of the title of the present paper, "Gene-environment interaction in the era of precision medicine", has been stolen from a recent paper in which bottlenecks of classical models of genetic effects and of their use in genetic mapping are discussed (Li et al., 2019). That paper echoes the message that techniques using conventional genetic models do often not provide insightful enough results and that, in particular, they provided hitherto no clear-cut evidence on whether disease etiologies are due to rare alleles with strong effects or to common alleles with weak effects. More to the point, that paper includes a simulation by means of which certain genetic models are shown not to be able to capture the complexity of the realistic underlying factors of a disease—particularly, involving epistatic effects (gene interactions, i.e., departures from the sum of the marginal contributions of the effects of the genes involved).

---
[1] Email address: jose.m.castro.alvarez@edu.xunta.gal

Further on, the aforementioned paper (Li et al., 2019) provides a probabilistic approach consisting on a Bayesian framework to hierarchically model gene-environment interaction, leading to a population-dependent index C called the genetic coefficient of the disease (at a population)—"a large C indicates large distinguishability of case genomes from control genomes". Then the paper illustrates the performance of the proposed methodology using a built-up example in which the disease susceptibility is by default very low (0.01) and it significantly increases due to either environmental (exposure) or genetic (risk allele) factors or both, to 0.4, 0.5 and 0.9, respectively. That case is hereafter referred to as the risk and exposure (RAE) case (see Table 1). With an exposure frequency of 0.24 and a frequency of the risk allele of 0.15, Li *et al.* (2019) report the genetic coefficient of the disease of the RAE case to be C=0.79.

**Gene-environment interaction within the framework of the classical models**

About half a dozen years earlier, Ma *et al.* (2012) had provided a model of gene-environment interaction based on the NOIA model of genetic effects (Álvarez-Castro and Carlborg, 2007), thus stemming from the classical models. In these models, the parameter $2\alpha$ can be used to reflect the "difference between the additive expectations of case genomes and control genomes", thus providing an alternative measure for Li *et al.*'s (2019) genetic coefficient of the disease, C. Assuming then Hardy-Weinberg proportions at the risk allele locus and equal risk of heterozygotes and homozygotes for the risk allele (since not explicitly specified otherwise in that paper), Ma *et al.*'s (2012) model can be used to compute a difference between the additive expectations of case genomes and control genomes of $2\alpha=0.85$ (or, to be more precise, $2\alpha_G=0.85$, using Ma *et al.*'s (2012) specific notation). The departure between this value and Li *et al.*'s (2019) genetic coefficient of the disease, C=0.79, could be due to the choices needed to be made above in relation with dominance and the Hardy-Weinberg proportions.

Besides the aforementioned statistical formulation of genetic effects, both NOIA (Álvarez-Castro and Carlborg, 2007) and Ma *et al.*'s (2012) extension of it to gene-environment interaction entail a so-called functional formulation. Whereas the statistical formulation is population-referenced and thus its parameters reflect properties of populations, the functional formulation is individual-referenced and thus its parameters reflect plane effects of substitutions from a reference class (a genotype at an environment) to the others. Applying that functional formulation from the default (non-exposed and non-risk) individual reference (0.01), the additive, dominance, environment, additive-by-environment and dominance-by-environment effects reflecting the aforementioned substitutions are 0.245, 0.245, 0,39, 0.005 and 0.005, respectively (see Table 2). Those values show that, although Li *et al.*'s (2019) RAE case entails both genetic and environmental effects, it can hardly be considered a gene-environment interaction case as intended, since the gene-environment interaction effects are extremely small relative to both the genetic and the environmental marginal contributions—the interaction effects actually lay about two orders of magnitude below the marginal effects.

Hitherto, it has been shown that relatively recent implementations of the classical models not only enable the analysis of the RAE case built up by Li *et al.* (2019) to illustrate their theoretical proposals but are also adequate to easily and precisely quantify basic properties of that case itself, which have apparently been missed by those authors. More generally speaking, theoretical developments stemming from the classical models are not always fairly acknowledged. To this regard it is at this point worth adding up that both NOIA (Álvarez-Castro and Carlborg, 2007) and Ma *et al.*'s (2012) extension of it to gene-environment interaction can properly deal with departures not only from complete dominance but also from Hardy-Weinberg proportions, which were assumed above only due to the absence of any explicit specifications of departures from those features.

Nevertheless, the general warning Li *et al.* (2019) post on the use of genetic models still holds—the current state-of-the-art of implementations of classical models of genetic effects, whether unfairly acknowledged or not, keeps on leaving room for further improvement. Indeed, the original NOIA proposal fails to properly account for nonrandom associations of marginal genotypic frequencies (*i.e.*, assumes linkage equilibrium between/among the loci involved) and Ma *et al.*'s (2012) inherits that limitation in what regards associations between genotypes and environments (*i.e.*, gene-environment correlations). Thus, those models shall hereafter be referred to as APNOIA (associations-pending NOIA) developments. Incidentally, it is imperative to overcome that limitation both because correlations between/among marginal frequencies may occur in populations and because they are in any case likely to achieve significant levels in the actual samples used in real data analyses.

**A further improved classical model of gene-environment interaction**

Opportunely, it is hereafter shown that the downsides of the APNOIA developments of gene-environment interaction can be fixed. Indeed, new mathematical developments for studying gene-environment interaction are provided right below in which gene-environment correlation is properly implemented. Since the resulting theoretical proposal overcomes the aforementioned associations-pending limitation, it shall be referred as ARNOIA (associations-resolved NOIA).

A biallelic locus A (with alleles $A_1$ and $A_2$) and two environmental instances ($E_1$ and $E_2$) of an environmental variable E are initially considered. This setting leads to six possible classes—combinations of genotypes and environments—and thus to six phenotypic expectations (*e.g.* six values of disease susceptibility). Those values are gathered in the column-vector of genotypic values, **G** = ($G_{ijk}$), where the subscripts indicate genotype $A_jA_k$ at environment $E_i$.

The genotypic values can be expressed in terms of genetic effects by means of regression model

$\mathbf{G} = \mathbf{N}_\mu\,\mu + \mathbf{N}_e\,\mathbf{e} + \mathbf{N}_\alpha\,\boldsymbol{\alpha} + \mathbf{N}_\delta\,\boldsymbol{\delta} + \mathbf{N}_{\alpha e}\,\boldsymbol{\alpha e} + \boldsymbol{\delta e}$, (expression 1)

in which the explanatory variables are the mean phenotype $\mu$, the environmental effect, $\mathbf{e} = \boldsymbol{\upsilon}_1 = (e_1, e_2)^T$ (where T stands for the transpose operation), the genetic additive effect, $\boldsymbol{\alpha} = \boldsymbol{\upsilon}_2 = (\alpha_1, \alpha_2)^T$, the dominance effect, $\boldsymbol{\delta} = \boldsymbol{\upsilon}_3 = (\delta_{11}, \delta_{12}, \delta_{22})^T$, and the additive-by-environment effect, $\boldsymbol{\alpha e} = \boldsymbol{\upsilon}_4 = (\alpha e_{11}, \alpha e_{12}, \alpha e_{21}, \alpha e_{22})^T$, and the residual term is the dominance-by-environment effect, $\boldsymbol{\delta e} = \boldsymbol{\eta}_4 = (\delta e_{ijk})$.

Let $\mathbf{1}^{(m)}$ be a column vector of length $m$ with all its scalars equal to 1, $\mathbf{I}^{(n)}$ an identity matrix of dimension $n$, $\mathbf{N} = \begin{pmatrix} 2 & 1 & 0 \\ 0 & 1 & 2 \end{pmatrix}^T$ and $\otimes$ the Kronecker product. Then, the design matrices in expression (1) can be expressed as

$\mathbf{N}_\mu = \mathbf{1}^{(6)}$, $\mathbf{N}_e = \mathbf{N}_1 = \mathbf{I}^{(2)} \otimes \mathbf{1}^{(3)}$, $\mathbf{N}_\alpha = \mathbf{N}_2 = \mathbf{1}^{(2)} \otimes \mathbf{N}$, $\mathbf{N}_\delta = \mathbf{N}_3 = \mathbf{1}^{(2)} \otimes \mathbf{I}^{(3)}$ and $\mathbf{N}_{\alpha e} = \mathbf{N}_4 = \mathbf{I}^{(2)} \otimes \mathbf{N}$. (expression 2)

Regression (1) with design matrices (2) is meant to be solved sequentially, as follows. Let the population frequencies be $p_{ijk}$ and let $\mathbf{P}$ be the diagonal matrix of those frequencies, $\mathbf{P}=\text{diag}(p_{ijk})$. Then, the mean phenotype is $\mu = \Sigma p_{ijk} G_{ijk}$, the mean-corrected vector of genotypic values is $\boldsymbol{\eta}_0 = \mathbf{G} - \mathbf{1}^{(6)}\mu$ and the expressions for the remaining explanatory variables and the residual term of regression (1) come from computing, for $l$=1 to 4,

$\boldsymbol{\upsilon}_l = \widetilde{\mathbf{H}}_l\,\boldsymbol{\eta}_{l-1}$ and $\boldsymbol{\eta}_l = \mathbf{M}_l\,\boldsymbol{\eta}_{l-1}$, (expression 3)

with $\widetilde{\mathbf{H}}_l = \left(\mathbf{N}_l^T \mathbf{P} \mathbf{N}_l\right)^{-1} \mathbf{N}_l^T \mathbf{P}$ and $\mathbf{M}_l = \mathbf{I}^{(6)} - \mathbf{N}_l \widetilde{\mathbf{H}}_l$.

In order to fully integrate the present theoretical proposal within the aforementioned NOIA framework (Álvarez-Castro and Carlborg, 2007), regression (1) has to be expressed in the form of a standardized statistical formulation. Such formulation is

$\mathbf{G} = \mathbf{S}\,\mathbf{E}$, (expression 4)

where $\mathbf{E} = (\mu, \alpha, \delta, e, \alpha e, \delta e)^T$ is the vector of genetic/environmental effects and $\mathbf{S}$ can be obtained via its inverse, $\mathbf{S}^{-1}$, whose rows are: the first one is $(p_{ijk})$, as it can be derived from $\mu = \Sigma p_{ijk} G_{ijk}$; the second one comes from $\alpha = (\alpha_2 - \alpha_1)$; the third one from $\delta = \delta_{12} - ((\delta_{11} + \delta_{22})/2)$; the fourth one from $e = e_2 - e_1$; the fifth one from $\alpha e = (\alpha e_{11} - \alpha e_{12} - \alpha e_{21} + \alpha e_{22})$ and the sixth one is $(½, -1, ½, -½, 1, -½)$.

Using previous extensions of classical models of genetic effects (Alvarez-Castro and Crujeiras, 2019; Álvarez-Castro and Yang, 2011), the ARNOIA regression framework for gene-environment interaction developed right above can be extended to several possibly multiallelic loci, arbitrary epistasis, arbitrary departures from linkage equilibrium and simultaneously to several environmental variables with multiple environmental instances, with nonrandom associations (*i.e.*, correlations) of environmental variables and of genotypes and environments. The details of such extensions are though out of the scope of this paper.

**How much of an improvement?**

For illustrating the advantage ARNOIA confers over the APNOIA shoulders it stands on (the ones of Álvarez-Castro and Carlborg, 2007; Ma et al., 2012), Figure 1A shows the additive, dominance and environmental effects (with dashed, dotted and solid lines, respectively) of the RAE case worked out above (see also Table 2). Ma *et al.*'s (2012) model provides—along the whole range of possible correlations between the risk allele and environmental exposure—the effects that fit to the random association scenario (gray horizontal lines), whereas ARNOIA (black lines) shows how those parameters actually change with negative (to the left of zero) and positive (to the right) risk-exposure correlations. Roughly, the effects decrease and increase with negative and positive correlations, respectively, although a slight decrease of the additive effect towards the maximum positive correlations and a bit of a more capricious behavior of the dominance effect for intermediate positive correlations can also be noticed. The thick vertical line marks the point of random association (*i.e.*, no correlation) where all values provided by APNOIA type Ma *et al.*'s (2012) model are correct and meet the ones provided by ARNOIA (particularly, by expression 4). The values of all the effects as computed at that point are provided in Table 2.

In view of Figure 1A, it could seem that settling for the relatively simpler APNOIA formulae (not accounting for nonrandom associations of genes and environments) by Ma *et al.* (2012) would not come with a high cost. Indeed, values that are correct for circumstances known beforehand (precisely, nonrandom associations) are retrieved regardless the nonrandom associations involved. However, that is but a mirage for such a restriction shall, on the one hand, compromise the flexibility of the model for making predictions, as illustrated further below, and, on the other hand, make the models to be less efficient in disclosing genetic architectures, as explained hereafter.

It is well known that interactions (of any kind, including gene-environment interaction or just dominance) may make lower level effects (like environmental effects or genetic additive effects) vanish under certain circumstances (possibly, given the genetic/environmental composition of an experimental sample). This is unfortunately not always properly taken into account. Such is the case, for instance, of a commendable review on models of gene-environment interaction in the context of plant breeding by Malosetti *et al.* (2013). Although the approaches there considered are not so genetic-effects grounded as NOIA (whether APNOIA or ARNOIA), they also found a sequential strategy like the one proposed above within expressions (1-3) to be most adequate. However, they overstepped the mark when specifically proposing a conditional sequential procedure, by claiming that "dominance effects should be tested conditioned on the additive effects present in the model." Since effects on the phenotype may (as further illustrated below) cancel out in average at the group of individuals under study because of the interactions in which they are involved, and thus become likely to be missed in mapping experiments, unveiling interactions becomes doubly imperative rather than something to subject to the condition of first having found their lower-order effects.

Thus, the theoretical genetic/environment models and the estimation strategies used must become as flexible and thorough as possible in order to address the difficulty of dealing with possibly masked effects. It is then necessary in the first place to opportunely implement such models with interaction effects, as thoroughly recalled by

Li *et al.* (2019). But since interactions are particularly elusive, it is also crucial to improve the flexibility of the models concerning the frequencies of the sample—specifically, concerning the departures of the frequencies of the sample from equilibrium situations like Hardy-Weinberg equilibrium, linkage equilibrium and random associations of genes and environments. Indeed, Ma *et al.* (2012) reasonably stress that their proposal implements—as an improvement from previous gene-environment interaction models—arbitrary departures from Hardy-Weinberg equilibrium.

In what follows, a case of actual gene-environment interaction is considered. It is a case of (genetic) risk to (environmental) exposure (thus referred to hereafter as RTE), where the risk allele increases disease susceptibility only when combined with exposure, hence actually interacting with the environment. Thus described, the interaction behaves as a switch—the environmental effect shall either be switched on (when carrying the risk allele) or turned off (otherwise). Table 1 provides the details of the RTE case. Table 2 further shows that the functional additive and dominance effects (*i.e.*, the marginal genetic effects) of this case from the reference of the individual default class (no genetic risk and no exposure) are zero, which actually is in accordance with the genetic risk being turned off in the absence of exposure. In Table 2 it is further shown that gene-interaction effects are large, which implies that the marginal genetic effects are not nil in absolute terms—they would show up from alternative references. In fact, still in Table 2, statistical (population-reference) marginal genetic effects are shown to be different from zero.

For a broader scope, Figure 1B shows all the genetic/environmental effects of the system as obtained using ARNOIA, with the marginal effects displayed as in Figure 1A and the gene-environment interaction effects in gray. The marginal genetic effects of the RTE case are small in the absence of gene environment correlation. Indeed, this case entails a visual example of a warning issued above since it illustrates that marginal effects approach zero as an occasional outcome (of a particular set of population frequencies), making it tricky to spot them in a mapping experiment. The trouble vanishes though as long as the (larger) gene-environment interaction effects are inspected (despite the apparent absence of marginal genetic effects) and disclosed. Note also that although the marginal genetic effects get closer to zero under certain negative correlations (towards the far left end of the graph), the additive-by-environment interaction effect increases accordingly. Thus, in any case, eventually out-of-reach marginal effects may be unveiled by diligently fishing interaction effects.

Overall, for properly detecting marginal (genetic and environmental) and interaction (gene-gene and gene-environment) effects (and, therefore, identify their corresponding loci and environmental variables) in mapping experiments it is essential that the genetic models entail not only any interactions between/among the effects themselves but also any departures from equilibrium genotype/environment frequencies, as Figure 1 shows ARNOIA to accomplish. Moreover, it is hereafter illustrated that the advantages of ARNOIA are crucial also for using detected genetic and environment underlying factors of traits in the formulation of predictions, particularly in the context of precision medicine.

**Predictions under diminishing exposure**

Figure 2A shows the genetic coefficient of the disease *sensu* Li *et al.* (2019) for the RAE case. As mentioned above, in the context of the developments stemming from the classical models of genetic effects such coefficient is given by the parameter $2\alpha$. On top of the variables already considered in Figure 1A, Figure 2A has one dimension added for enabling predictions in the context of a hypothetical decrease of the environmental exposure. The black solid line in Figure 2A marks random association and shows that the genetic coefficient of the disease is simply not affected by decreasing the exposure frequency in the population. This is as expected under lack of interplay between gene an environment (*i.e.*, no interaction and to correlation). Indeed, although the trait is subject to both genetic and environmental influence, as long as there is no (or very little) interplay between them, the genetic parameter remains virtually constant in the face of variations in the environmental exposure.

However, as already shown above in relation with Figure 1 (where the additive genetic effect, $\alpha$, was shown instead of the slightly different genetic coefficient of the disease, $2\alpha$), such interplay may come not only by means of gene-environment interaction but also through gene-environment correlation. Thus and so, whereas the genetic coefficient of the disease remains constant in Figure 2A against diminishing exposure in the absence of significant gene-environment interaction, it is in point of fact affected by risk-exposure correlations. In particular, the genetic coefficient of the disease decreases with negative associations, as the surface to the left of the black line shows, and it increases for positive associations up to a maximum followed by a slight decrease, to the right of the black line (likewise the additive effect, $\alpha$, in Figure 1A). Note also that the range of risk-exposure associations narrows down as the exposure frequency approaches zero, which explains the tip of the surface at the end of the black line.

In Figure 2B, the RTE case of Figure 1B is resumed and further extended in a way analogous to Figure 2A from Figure 1A. As Figure 2B shows, for the RTE case the genetic coefficient of the disease approaches zero for decreasing values of exposure under random association of risk and environment (decreasing black line). That coefficient also decreases—also down to zero—for decreasing (increasingly negative) association between the risk allele and environmental exposure, as the left tip of the surface shows. In plain language, the figure shows that the problem of increased disease susceptibility of the carriers of the risk allele may be equally reduced (and eventually removed) either by reducing exposure for the whole population or by restricting the access to the exposed environment only for the risk population, or even trough any intermediate alternative (any reduction of the exposure in the population biased towards the carriers of risk alleles). The optimal management would then depend just upon the reluctance of the average individual to avoid the exposed environment (or even the actual feasibility of bringing the whole population out of it) and the cost of tests to detect the risk allele, which would enable personalized warnings.

Overall, the two cases considered in Figure 2 deal with rather singular—either virtually absent, RAE, or switch-type, RTE—instances of gene-environment interaction, for which some predictions would be feasible even without mathematical modeling. The results obtained using ARNOIA not only reassuringly agree with the conceptually attainable predictions but also further illustrate how to precisely quantify any desired genetic/environmental parameter, which can hereinafter be applied to more complex real cases of interest undergoing less intuitive behaviors.

**Road work ahead**

Because affordable data is an ever-changing variable, it is sensible to assume that, likewise, theoretical models required in the analyses shall need to keep on being worked out every now and then. In this context, it is as essential to make the best possible use of the models available at a particular time-spot as it is to point out in which way they are at that time imposing limitations in the analyses. For instance, as Li *et al.* (2019) have claimed, it is fundamental to interiorize the importance of epistasis as, *e.g.*, Huang *et al.* (2012) showed. But then Li *et al.* (2019) are also implicitly acknowledging that "model insufficiency" was not so severe as to preclude the crucial importance of epistasis in genetic architectures to be inferred. More to the point, an elusive implementation of models of genetic interactions (epistasis)—even claimed to be beyond reach—has recently been developed, enabling genetic interaction and genotype frequencies correlation (linkage disequilibrium) to be disentangled (Alvarez-Castro and Crujeiras, 2019). Analogous achievements are provided in this communication in what regards gene-environment interaction and gene-environment correlation, as discussed hereafter.

In what is currently understood as the, at least relative, "failure of GWAS" (Teperino, 2020b), gene-environment interaction is also pointed out as a key factor. Indeed, the importance of gene-environment interaction in human health has been stressed in relation with a broad spectrum of disorders ranging from obesity and other metabolic disorders through autoimmune diseases to cancer (e.g. Cust, 2020; Teperino, 2020a). What is then necessary in order to move forward? Concerning the theoretical models involved in data analyses, developments enabling gene-environment interaction and gene-environment correlation to be properly disentangled have especially been demanded. As a recent example within the field of precision medicine, particularly in the context of mental health, Assary *et al.* (2020) have advocated that "Identifying which form of gene-environment interplay contributes to a particular disorder or behavior is absolutely crucial in order to select suitable intervention efforts" because theoretical developments that enable a joint analysis of both phenomena are needed in particular for "ensuring that the outcomes of one do not bias the effects of the other".

This paper meets that demand and it does so by providing a theoretical framework that, as mentioned above, simultaneously addresses many other genetic facts of relevance (like epistasis, multiple alleles and departures from genetic equilibria like Hardy-Weinberg equilibrium and linkage equilibrium). Thus, more to the point, the ARNOIA model here provided illustrates the possibilities of mathematical developments stemming from the classical models of genetic effects in what regards their potential to keep on being improved and address eventual demands to come. In other words, no evidence seems to support that the machinery here proven useful to fix inconvenient potholes of the classical models road could be inadequate for similar purposes in the future.

The previous is however not to say that alternative roads should never be built. It looks sensible in any case to assume that a new road will consume significant resources before providing benefits comparable to the already existing ones, especially in what regards the wealth of experience amassed in the use of them. Therefore, it would be reasonable to first thoroughly inspect the possibilities of the existing roads to be fixed and as well to guarantee the added value the new road is intended to bring. On top of that, it would also make perfect sense to assume that the new road would only provide its best service when adequately connected with the previous road network. Whenever developed along these lines, alternative perspectives in genetic modeling could aim to open doors to novel analyses and/or double check the already existing ones and thus enrich the application of mathematical models in precision medicine.

As a final remark, it would be regrettable in the context of the current COVID-19 pandemic not to devote at least a few lines of this paper to the application of ARNOIA to such a global threaten. The strong link between gene-environment interaction and epidemiology has already been underscored in relation with the COVID-19 pandemic (Rodriguez-Morales et al., 2020). It thus becomes worth making it crystal clear in the first place that ARNOIA is certainly suitable not only for the trait considered in the examples above (disease susceptibility), but also for additional traits of relevance within the field of epidemiology, including for instance mortality caused by a disease. It is also particularly useful to notice here that although virulence variability is underlain by mutations (and thus conceptually related to genetics), ARNOIA may naturally integrate the virus mutations simply as an environmental variable, since that is how they are perceived from the perspective of the susceptible individuals—the genetic component of the model. Bearing that in mind, it is easier to perceive why it is crucial, for the study of COVID-19 as well, that ARNOIA considers together (but disentangled) gene-environment interaction and gene-environment correlation. Indeed, the various geographical regions affected by the disease do not only undergo different proportions of virus strains (environmental component) but also different genetic backgrounds of the susceptible individuals (genetic component), thus setting a gene-environment correlation scenario in which gene-environment interaction needs to be properly understood.

## Acknowledgments

The author acknowledges insightful comments by Ania Pino-Querido, which improved the final form of this paper.

## Declaration of interests

The author declares no competing interests.

**Figure legends**

**Figure 1:** Genetic and environmental effects of disease susceptibility influenced by a risk allele and environmental exposure, for the whole range of possible correlations (including negative and positive associations) of the risk allele and environmental exposure. The risk allele frequency is 0.15, with genotypic frequencies under Hardy-Weinberg equilibrium, and the environmental exposure frequency is 0.24. The RAE case (see Table 1) is shown in Figure 1A, where APNOIA and ARNOIA, shown with gray and black lines, respectively, are compared. The thick black solid vertical line marks the case of random association (*i.e.*, no correlation) between risk and exposure. The genetic effects obtained with ARNOIA for the RTE case (see Table 1) are shown in Figure 1B. The marginal genetic and environmental effects are shown with the same black lines as in Figure 1A and gray lines are used here for the interaction effects.

**Figure 2:** Genetic coefficient of the disease obtained with ARNOIA for the RAE and RTE cases. The details are as in Figure 1, plus an extra dimension for decreasing values (down to zero) of exposure. The values of the vertical axis range from 0 to 1.20. The thick black line marks the absence of correlation between risk allele and environmental exposure, which are the ones APNOIA would provide for the whole range of correlations between the risk allele and environmental exposure.

**Table 1:** Phenotypes (disease susceptibility) of the four individual classes (risk allele carriers and non-carriers under exposed and non-exposed environments), for the two cases considered in the text—the case taken from Li *et al.* (2019), here called the risk and exposure (RAE) case and the genetic risk to exposure (RTE) case. Complete dominance of the risk allele is assumed so that homozygotes for the risk allele and heterozygotes are equally susceptible to the disease.

| Case | Environment | Genetics | |
| --- | --- | --- | --- |
| | | Default | Risk |
| RAE | Default | 0.01 | 0.5 |
| | Exposed | 0.4 | 0.9 |
| RTE | Default | 0.01 | 0.01 |
| | Exposed | 0.4 | 0.9 |

**Table 2:** Genetic/environmental effects of the two cases, RAE and RTE, considered in the text and detailed in Table 1. AEI and DEI stand for additive-environment interaction and dominance-environment interaction, respectively. For each of the two cases, this table shows both functional (*i.e.* individual-referenced) effects from the reference of the default (non-risk and non-exposed class, with a disease susceptibility of 0.01) and statistical (*i.e.* population-referenced) effects from the reference of the average phenotype. The reference population of the statistical effects has a frequency of exposure of 0.24, a frequency of the risk allele of 0.15, Hardy-Weinberg proportions and random associations of (*i.e.*, absence of correlation between) genotypes and environments. By virtue of the latest, the results here reported may be equally obtained using APNOIA and ARNOIA (ultimately from expression 4), as explained in the text.

| Case | Reference | Genetic/environmental effects | | | | |
| --- | --- | --- | --- | --- | --- | --- |
| | | Additive | Dominance | Environment | AEI | DEI |
| RAE | 0.01 | 0.245 | 0.245 | 0.39 | 0.005 | 0.005 |
| | 0.245 | 0.393 | 0.246 | 0.394 | 0.008 | 0.005 |
| RTE | 0.01 | 0 | 0 | 0.39 | 0.25 | 0.25 |
| | 0.081 | 0.096 | 0.06 | 0.57 | 0.4 | 0.25 |

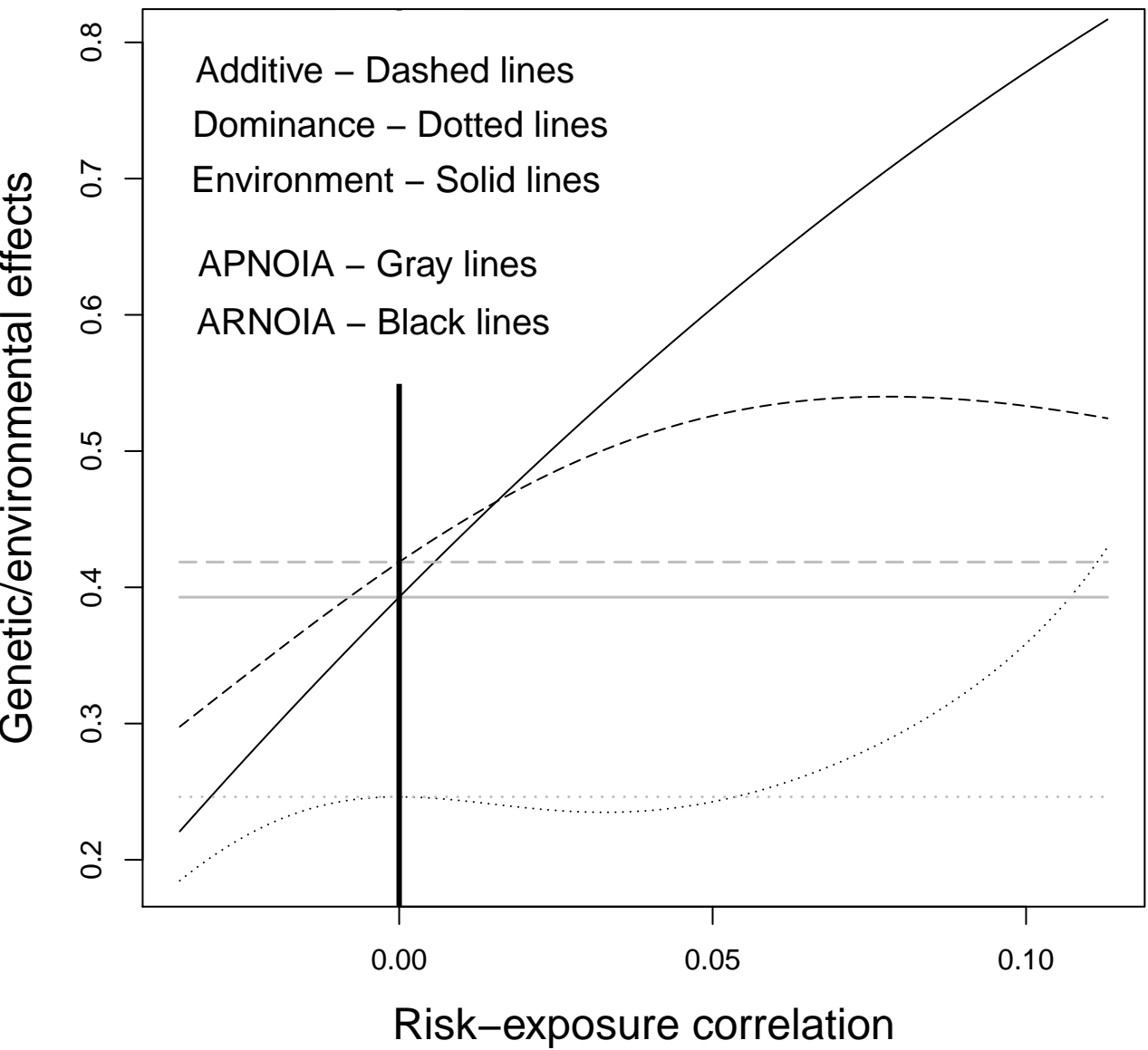

**1A – Risk and exposure (RAE)**

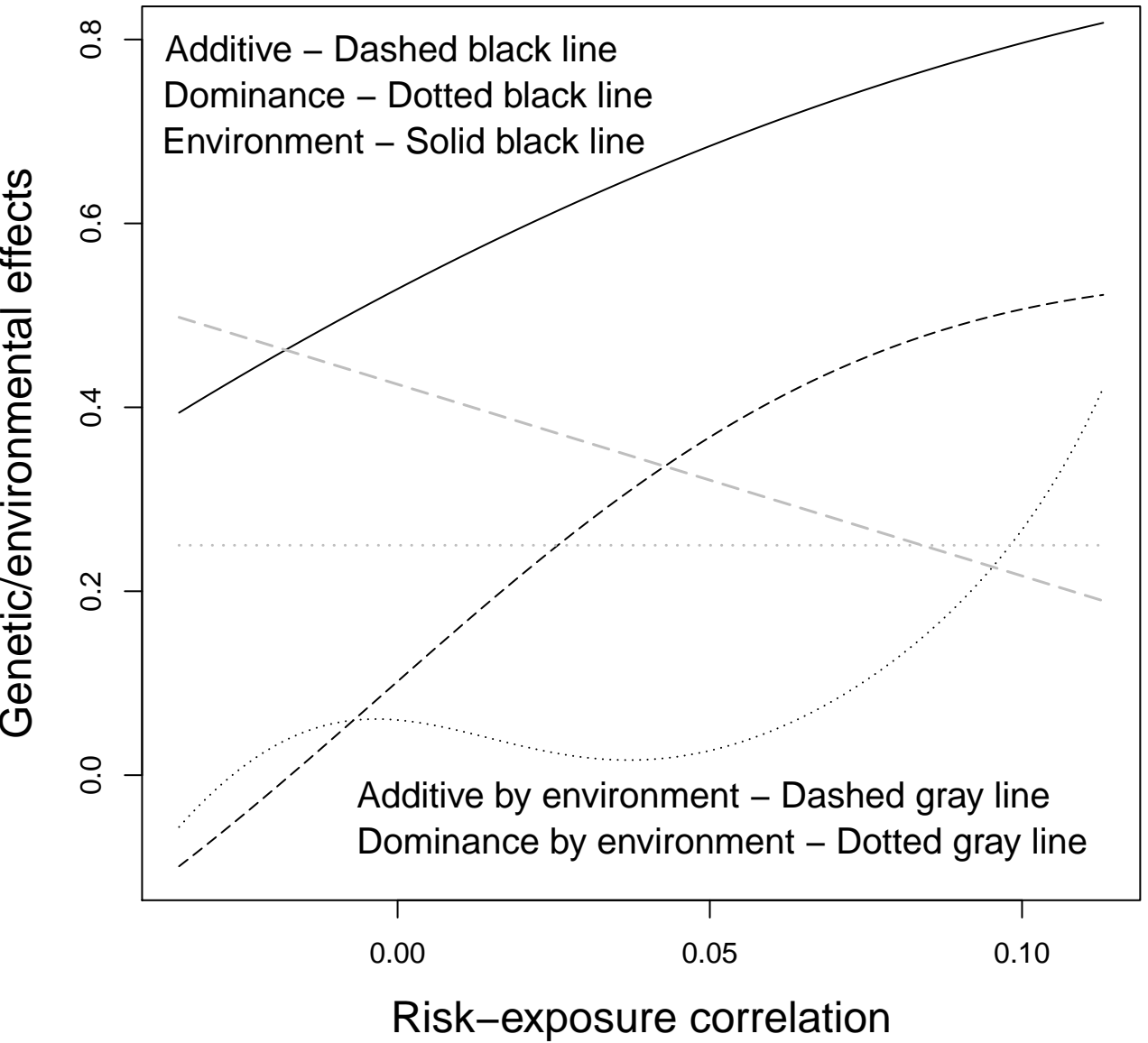

**1B – Risk to exposure (RTE)**

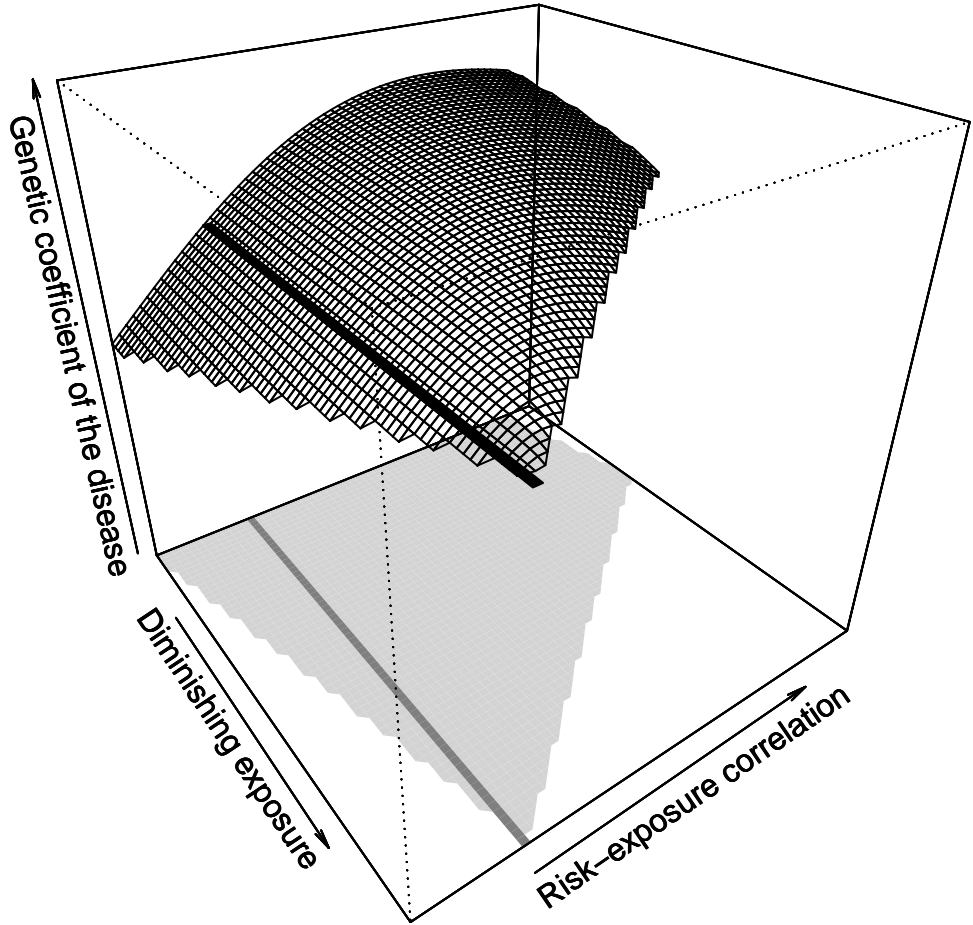

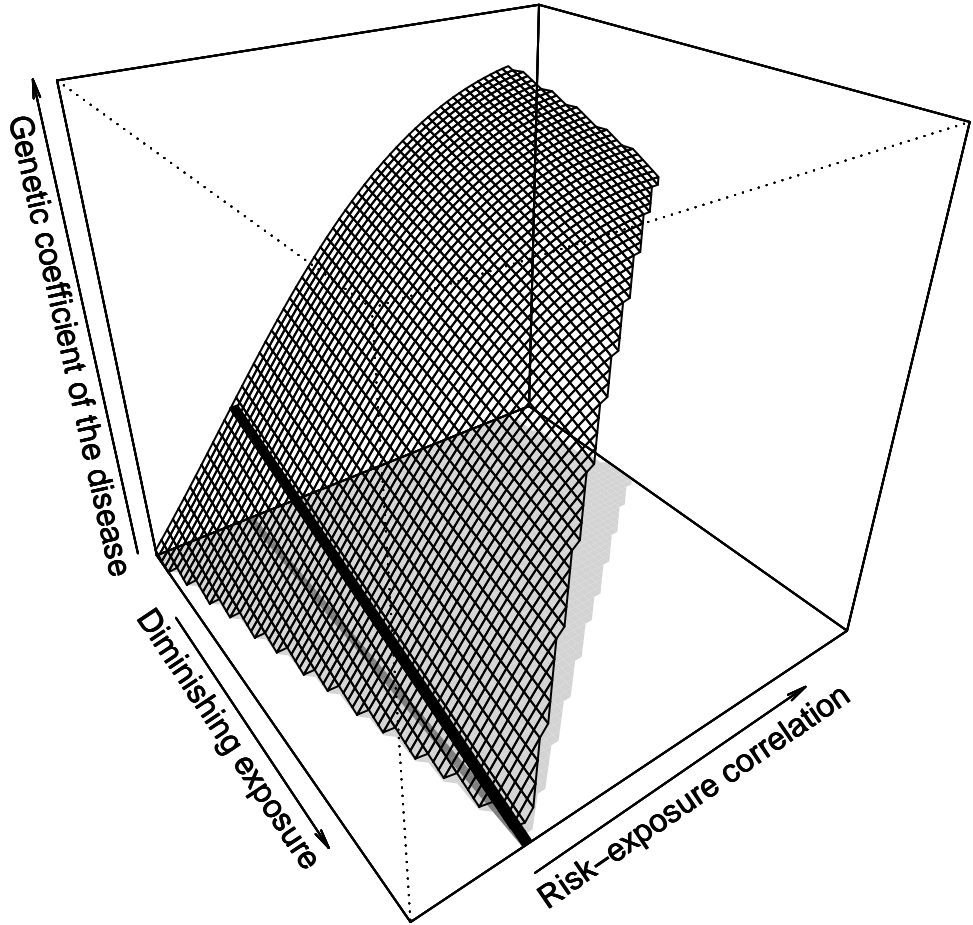